# Pan-cancer gene set discovery via scRNA-seq for optimal deep learning based downstream tasks

Jong Hyun Kim[1,+] and Jongseong Jang[1,*]

[1]LG AI Research
[*]j.jang@lgresearch.ai

## ABSTRACT

The application of machine learning to transcriptomics data has led to significant advances in cancer research. However, the high dimensionality and complexity of RNA sequencing (RNA-seq) data pose significant challenges in pan-cancer studies. This study hypothesizes that gene sets derived from single-cell RNA sequencing (scRNA-seq) data will outperform those selected using bulk RNA-seq in pan-cancer downstream tasks. We analyzed scRNA-seq data from 181 tumor biopsies across 13 cancer types. High-dimensional weighted gene co-expression network analysis (hdWGCNA) was performed to identify relevant gene sets, which were further refined using XGBoost for feature selection. These gene sets were applied to downstream tasks using TCGA pan-cancer RNA-seq data and compared to six reference gene sets and oncogenes from OncoKB evaluated with deep learning models, including multilayer perceptrons (MLPs) and graph neural networks (GNNs). The XGBoost-refined hdWGCNA gene set demonstrated higher performance in most tasks, including tumor mutation burden assessment, microsatellite instability classification, mutation prediction, cancer subtyping, and grading. In particular, genes such as *DPM1*, *BAD*, and *FKBP4* emerged as important pan-cancer biomarkers, with *DPM1* consistently significant across tasks. This study presents a robust approach for feature selection in cancer genomics by integrating scRNA-seq data and advanced analysis techniques, offering a promising avenue for improving predictive accuracy in cancer research.

## Introduction

Recent advancements in oncology have harnessed the potential of machine learning technologies to enhance the analysis of transcriptomic data across various cancer types[1,2]. By integrating transcriptomic data with advanced machine learning models, researchers were able to improve the accuracy of complex downstream cancer tasks such as mutation prediction, cancer subtyping and survival prediction[3–5]. In particular, models like gradient boosting and neural network architectures, including multilayer perceptrons (MLPs) and graph neural networks (GNNs), have demonstrated their capability in extracting meaningful patterns from complex genomic datasets[6,7]. These advancements facilitate accurate predictions, offering new clinical perspectives and elevating the potential of pan-cancer studies, ultimately improving the development of targeted therapies and personalized treatment strategies.

RNA sequencing (RNA-seq) is a powerful technique in cancer research, providing profound insights into the genetic landscape of tumors. However, the high dimensionality and complexity of RNA-seq data present significant analytical challenges, requiring robust feature selection to identify the most relevant genes[8]. Effective feature selection is crucial not only for accurate biological interpretation but also for enhancing the performance of predictive models, particularly those employing deep learning techniques[9]. Various feature selection methods have been developed for RNA-seq data in cancer studies, but no generalized method fits all applications, leading to diverse approaches. For instance, Chen et al.[10], utilized pathway-based gene sets relevant to cancer from the MsigDB hallmark database. While Elbashir et al.[11], focused on differentially expressed genes, and Silva et al.[12], selected genes with the highest variance across samples. This diversity underscores the lack of standardization in feature selection, making it challenging to identify consistent patterns across studies. This challenge is particularly notable in pan-cancer studies, where identifying commonalities across cancer types is essential.

Single-cell RNA sequencing (scRNA-seq) has revolutionized our understanding of the tumor microenvironment by revealing the cellular diversity and dynamics often obscured in bulk RNA sequencing data[13]. This technology is particularly valuable in pan-cancer studies, offering a detailed perspective on immune cell diversity and behavior within tumors, which significantly influences cancer progression and treatment response[14]. Recent advances in scRNA-seq have illuminated specific cellular mechanisms, enhancing our understanding of the immune landscape across multiple cancers[15–17]. By revealing these cellular behaviors, scRNA-seq facilitates comprehensive biomarker discovery and predictive modeling, providing a foundation for future studies to explore the impact of cellular details on broader cancer dynamics. Furthermore, scRNA-seq enables high-resolution identification of key genetic signatures in a variety of cancers that were not previously possible. This allows researchers to select features based on high-resolution datasets, potentially shifting from a feature selection approach limited by bulk data to one enriched with detailed, cellular insights. These advancements not only improve our understanding of gene function in different cancer types but also enhance the predictive capabilities of computational models used in oncology.

In this study, we aim to address the challenge of optimal feature selection in pan-cancer research by leveraging scRNA-seq data. We hypothesize that gene sets derived from scRNA-seq data will outperform conventional bulk RNA-seq-derived gene sets in predictive modeling and downstream cancer tasks. This hypothesis was evaluated through a comprehensive analysis of scRNA-seq data from 181 tumor biopsies across 13 different cancer types. We applied high-dimensional weighted gene co-expression network analysis (hdWGCNA) to identify gene sets of high relevance. These gene sets were then refined using XGBoost to compute feature importance, allowing us to select the most significant features for each downstream task. We used TCGA pan-cancer RNA-seq data to perform these downstream tasks, using the selected gene sets from the scRNA-seq data. To validate our approach, we compared our gene sets to six reference gene sets and oncogene sets from OncoKB, an open pan-cancer database. The performance of these gene sets was evaluated using two different deep learning models, and our study found that XGBoost-refined hdWGCNA gene sets had the highest predictive accuracy in various cancer-related tasks. This study provides a robust framework for optimizing feature selection in cancer genomics, helping to identify the most effective gene sets for deep learning models.

## Results

### Overview of study

Our study workflow, as illustrated in Fig.1, we conducted a comprehensive analysis of scRNA-seq data from 181 tumor biopsy samples, encompassing 13 distinct cancer types (Fig.1a). This dataset includes a representation of 10 tumor types, providing a broad spectrum for analysis as indicated by the proportional distribution of cancer types (Fig.1b). From the scRNA-seq data, we processed hdWGCNA to generate a detailed gene co-expression network (Fig. 1c, d). This analysis allowed us to identify clusters of co-expressed genes that share biological functions. These co-expressed genes were subsequently used as gene sets for further downstream tasks. We focused on selecting genes that form part of the co-expression network, filtering out those that do not form a gene-gene network. These selected genes were then applied to feature selection processes of RNA-seq data for downstream modeling tasks, as shown in Figure 1e (Supplementary Table 1).

In addition, we used XGBoost to further refine these features by selecting the most significant features for each specific downstream task, ensuring that our analysis focused on the most relevant and meaningful genes. These downstream tasks, conducted using TCGA pan-cancer RNA-seq data, included tumor mutation burden (TMB) assessment, microsatellite stability (MSI) classification, mutation prediction, cancer subtyping, and grading. Our approach effectively integrates comprehensive single-cell data with advanced machine learning techniques to improve predictive modeling and provide insights critical for the development of targeted therapies and cancer diagnosis.

Our approach of combining high-resolution scRNA-seq data with advanced analytical techniques not only refines the selection of biomarkers but also enhances the accuracy of computational models. These results highlight the importance of advanced feature selection methods combined with powerful analytical tools in advancing cancer research.

### High-dimensional co-expression network through scRNA-seq

We generated a scRNA-seq dataset from 181 tumor biopsy samples, including 87,659 genes and 317,111 cells[18]. This dataset characterizes distinct immune cell subsets, providing a detailed view of the tumor microenvironment. UMAP embeddings (Fig. 2a) visualize the cellular diversity and distribution of 25 immune cell types across 181 samples.

To prepare for the hdWGCNA, we first filtered the genes to include only those expressed in more than 5% of cells. This process resulted in identifying 6,617 common genes across all cancer types, thereby reducing noise and ensuring the analysis focused on genes with consistent expression levels. Using the dynamic tree-cut method, we constructed co-expression networks, identifying hierarchical gene clusters and filtering out genes that did not contribute to these co-expression networks. This approach resulted in 11 modules, visualized in the gene network (Fig. 2b), comprising a total of 1,857 genes. Each module is represented by the top 25 hub genes with the highest interconnectivity, ranked by kME (Fig.2c). The genes constituting each module are listed in Supplementary Table 2.

To assess the biological significance of the identified modules, we conducted a gene ontology (GO) pathway enrichment analysis. The heatmap in Fig. 2d summarizes the enrichment results, highlighting the key biological processes associated with each module. Module 1 is involved in immune and apoptotic processes, such as the humoral immune response and apoptosis, which are important for immune suppression and tumor cell elimination in various cancer types. Module 2 focuses on mitochondrial energy metabolism, including cellular respiration and oxidative phosphorylation, which are essential to maintain the high metabolic demands of rapidly growing tumor cells in a pan-cancer context. Module 4 is involved in ribosome biogenesis and protein synthesis, including translation and gene expression, which are fundamental to the persistent growth and proliferation of cancer cells. Module 5 focuses on protein folding and stability, such as the cellular response to unfolded proteins and protein stabilization. Module 6 involves vascular and immune cell migration, such as endothelial cell movement and actin filament formation, which are crucial for tumor angiogenesis and metastasis in pan-cancer progression. Module 7 focuses on cell adhesion and phagocytosis, involving leukocyte adhesion and plasma membrane invagination, highlighting the role of the tumor microenvironment in cancer immunity and inflammation. Module 8 is linked to T cell activation and cytokine responses, including T cell activation pathways and interleukin signaling, emphasizing the importance of adaptive immunity in recognizing and eliminating tumors across various cancers. Module 9 covers vesicle transport, such as Golgi vesicle transport and endoplasmic reticulum to Golgi transport, vital for protein trafficking and secretion in cancer cells. Module 10 involves stress response and growth factor signaling, including responses to fibroblast growth factors and interferon-beta, reflecting the adaptive mechanisms cancer cells use to survive in hostile environments. Finally, module 11 is associated with transcriptional regulation and RNA processing, including mRNA splicing and RNA processing, essential for gene expression regulation and the adaptability of cancer cells to different stresses and therapeutic interventions.

**Comparative performance analysis of gene sets for downstream tasks**

After identifying gene modules through hdWGCNA, we applied feature selection using XGBoost to refine these modules for specific downstream tasks (Supplementary Table 3). We evaluated the importance of each feature. We retained genes with a feature importance score higher than 0.001 for each task. These tasks included TMB assessment, MSI classification, mutation prediction for *TP53*, *EGFR*, and *KRAS* genes, cancer subtyping, and grading. For validation, we performed a comparative analysis using our refined gene sets along with six reference gene sets and the OncoKB database gene set[19–25] (Supplementary Table 4 and 5). We applied a stratified 5-fold cross-validation strategy to ensure a robust performance evaluation. Results from each fold were averaged and bootstrapped to obtain 95% confidence intervals (CI) for the mean area under the receiver operating characteristic curve (AUROC). The performance of our gene sets was evaluated using two different models: MLP and GNN (see more details in the method).

Overall, the XGBoost-refined hdWGCNA gene sets outperformed the reference gene sets in downstream tasks, except for the breast invasive carcinoma (BRCA)-*TP53* mutation prediction task (Table 1).

The results are as follows:

> **Tumor mutation burden (TMB)**: For the TMB assessment, both the MLP and GNN models demonstrated the highest performance. Specifically, for lung adenocarcinoma (LUAD), the average AUROC was 0.791 (95% CI: 0.782-0.800) with the MLP model and 0.764 (95% CI: 0.754-0.773) with the GNN model. In the case of lung squamous cell carcinoma (LUSC), the MLP model achieved an average AUROC of 0.647 (95% CI: 0.636-0.659), while the GNN model achieved 0.604 (95% CI: 0.592-0.616). For skin cutaneous melanoma (SKCM), the average AUROC was 0.772 (95% CI: 0.762-0.782) with the MLP model and 0.747 (95% CI: 0.736-0.758) with the GNN model. Lastly, for colorectal cancer (CRC), the average AUROC was 0.826 (95% CI: 0.818-0.835) with the MLP model and 0.808 (95% CI: 0.799-0.817) with the GNN model.
>
> **Microsatellite instability (MSI)**: For stomach adenocarcinoma (STAD), the average AUROC was 0.990 (95% CI: 0.988-0.991) with the MLP model, while the OncoKB gene set performed best with the GNN model, achieving an AUROC of 0.982 (95% CI: 0.979-0.985). For CRC, the average AUROC was 0.931 (95% CI: 0.925-0.937) with the MLP model and 0.936 (95% CI: 0.930-0.942) with the GNN model, both outperforming the reference gene sets.
>
> **Mutation prediction (MUT)**: In mutation prediction tasks, the OncoKB gene set achieved the highest performance for BRCA-*TP53*, with an average AUROC of 0.951 (95% CI: 0.949-0.954) using the MLP model, and 0.935 (95%

CI: 0.932-0.938) with the GNN model. For LUAD-*TP53*, the average AUROC was 0.869 (95% CI: 0.862-0.876) with the MLP model and 0.872 (95% CI: 0.866-0.880) with the GNN model. Similarly, for LUAD-*EGFR*, the average AUROC was 0.868 (95% CI: 0.858-0.878) with the MLP model and 0.865 (95% CI: 0.855-0.875) with the GNN model. For LUAD-*KRAS*, the average AUROC was 0.845 (95% CI: 0.835-0.854) with the MLP model and 0.792 (95% CI: 0.782-0.801) with the GNN model. In the case of pancreatic adenocarcinoma (PAAD)-*KRAS*, the average AUROC was 0.904 (95% CI: 0.894-0.915) with the MLP model and 0.892 (95% CI: 0.879-0.904) with the GNN model. For STAD-*TP53*, both models achieved an average AUROC of 0.790 (95% CI: 0.779-0.800) with the MLP model and 0.790 (95% CI: 0.780-0.799) with the GNN model.

**Grading (GRAD)**: For cancer grading in prostate adenocarcinoma (PRAD), the average AUROC was 0.779 (95% CI: 0.771-0.787) with the MLP model, showing the highest performance among all gene sets. The GNN model had a slightly lower AUROC of 0.695 (95% CI: 0.686-0.704).

Despite these successes, the XGBoost-refined hdWGCNA gene sets did not consistently outperform the reference sets in all tasks. In MSI classification for STAD, the GNN model performed better with the OncoKB gene set, achieving an AUROC that was only 0.003 higher than that of the XGBoost-refined hdWGCNA gene set. Similarly, for BRCA-*TP53* mutation prediction, the OncoKB gene set performed best, but the differences were minimal. The MLP model had an AUROC difference of only 0.006 and the GNN model had an AUROC difference of only 0.001. Furthermore, we conducted subtyping classification tasks. However, it was observed that the performance metrics for all the reference gene sets were already remarkably high. Given the minimal differences in performance metrics, we have presented these results in Supplementary Table 6.

In summary, while the XGBoost-refined hdWGCNA gene sets demonstrated robust performance across many tasks, the performance differences compared to the reference sets were minimal. These results highlight the effectiveness of our approach in identifying optimal gene sets for cancer downstream tasks.

## Biological significance of XGBoost-refined hdWGCNA gene sets

To evaluate the biological relevance of the XGBoost-refined hdWGCNA gene sets, we analyzed the feature importance of selected genes within each downstream task. For each task, we identified the top 10 genes with the highest feature importance scores (Fig. 3a-d). Many of these genes are known to be associated with cancer biology, supporting their importance and potential use in the context of our study.[26–38].

In the context of *TP53* mutation prediction in BRCA, *CIRBP*, which is crucial for DNA repair and cell proliferation, has been identified as playing a pivotal role in triple-negative breast[26]. Similarly, *BTF3* has been reported to be up-regulated in individuals with *TP53* mutations across various cancer types[27]. In LUAD, *KPNA2*, a nuclear export protein essential for tumor formation, has been recognized for its significant role in human tumors[28]. Additionally, mutations in *CDKN2A* have been associated with *TP53* mutations, underscoring its relevance in this context[29]. For *TP53* mutation prediction in STAD, *DDB2*, closely linked to the p53 pathway, acts as a key regulator of p21 Waf1/Cip1 following DNA damage or induction by p53[30]. In the LUAD-*EGFR* mutation prediction task, *GGA2* was identified as a critical gene due to its interaction with *EGFR*, which increases *EGFR* protein levels and modulates its degradation[31]. For *KRAS* mutation prediction in LUAD, *RHOB*, known to be down-regulated in non-small-cell lung cancer (NSCLC), is directly related to lung cancer progression[32]. In PAAD, *S100A11*, a well-known pancreatic tissue marker, was identified as crucial due to its association with tumor stage, drug resistance, and shorter overall survival[33]. *YWHAZ*, also identified in PAAD, is recognized as a promising novel treatment target due to its role in promoting cell proliferation, migration, and invasion[34]. For MSI classification tasks, *RPL22L1* was significant in MSI-STAD due to its up-regulation in high-MSI groups across various cancer types[35]. In MSI-CRC, *PMEPA1* was notable for its association with tumor immunity in pan-cancer studies[36]. In TMB tasks, *PSMB9* was identified as a strong predictor of immune response in melanoma patients, significantly influencing their response to checkpoint therapies[37]. In TMB-CRC, *CCL5* was noted as a prognostic biomarker linked to the efficacy of immunotherapeutic interventions in cancers[38].

To identify the overlap of selected genes across downstream tasks, we visualized them using upset plots (Figure 3e). These plots highlight genes that consistently ranked high in feature importance across multiple tasks. In particular, *DPM1* was significant in all 13 tasks, highlighting its critical role in pan-cancer studies. *DPM1* has previously been identified as a potential prognostic tumor marker in hepatocellular carcinoma (HCC), further supporting its importance in cancer biology[39]. In addition, *BAD* was ranked high in all downstream tasks except the tumor mutation burden task in CRC, indicating its broad involvement in multiple cancer pathways. *FKBP4*, which appeared in 10 tasks, was associated with prognostic and immunological roles in several cancers[40]. *DUSP4*, associated with treatment resistance in breast cancer was presented[41]. Similarly, *LASP1* which is involved in metastasis in pancreatic cancer, was present in 9 tasks[42]. The overlap of these genes suggests that they are key players in common mechanisms across cancer types, which suggests that they are promising targets for further pan-cancer studies. Our findings confirm the relevance of these genes as individual cancer markers and highlight their potential as pan-cancer biomarkers.

## Discussion

This study demonstrated the significant potential of integrating scRNA-seq data with advanced feature selection methods to enhance predictive modeling in cancer genomics. By utilizing hdWGCNA and refining the results with XGBoost, we derived gene sets from scRNA-seq data that showed improved performance in various pan-cancer downstream tasks compared to gene sets derived from conventional bulk RNA-seq data. The high-resolution single-cell data captured intricate cellular heterogeneity and dynamic interactions within the tumor microenvironment, enhancing the biological interpretability of our models. Additionally, XGBoost refinement allowed for the selection of the most significant genes tailored to specific tasks, improving model accuracy and robustness. Many identified genes are well-known in cancer studies, validating the biological relevance and robustness of our approach. Furthermore, we identified additional genes with potential as pan-cancer biomarkers, representing promising targets for further investigation. Our findings support the integration of scRNA-seq data with advanced analytical techniques and contribute to the broader understanding of cancer biology by highlighting key genetic drivers.

Traditional feature selection approaches for RNA-seq data, such as those utilizing MsigDB or OncoKB, typically rely on predefined biological pathways and cancer biology-based gene sets[25,43]. These databases have demonstrated high performance in various studies due to their strong biological basis and have been highlighted as useful for capturing relevant pathways and processes. However, the complexity of the tumor microenvironment is often not captured by conventional methods that do not use such databases. For example, Mohammed et al. used LASSO regression to screen 173 genes with 10-fold cross-validation[11]. Similarly, Chen, Joe W., et al. selected the top 500 differentially expressed genes based on the lowest p-value[22]. Silva., et al have selected genes with the highest variance across samples[44]. While these methods are statistically robust, they are based on bulk RNA-seq data and often fail to account for the complex cellular interactions and heterogeneity within the tumor microenvironment. This can result in the loss of important cellular context and overlook genes that play crucial roles in small cell populations within the tumor. Additionally, the features curated by these methods are typically not in a pan-cancer context, limiting their applicability in identifying general biomarkers across multiple cancer types.

In contrast, our hdWGCNA approach revealed 11 distinct gene modules, each linked to biological processes such as immune response, mitochondrial metabolism, and protein synthesis. By using scRNA-seq data, we captured the cellular diversity and dynamics within tumors that are often missed in bulk RNA-seq data, thereby identifying more biologically relevant gene sets and increasing the validity of our results. Additionally, our refined gene sets generally outperformed reference gene sets in most tasks, demonstrating their robustness and potential for broader applications. This indicates that our approach not only improves predictive accuracy but also provides a more comprehensive understanding of the genetic underpinnings of various cancers. The success of our gene sets in these tasks underscores their potential utility in multi-modal research, where integrating diverse data types could further enhance predictive models.

In particular, our study highlights the potential of several genes, such as *DPM1*, *BAD*, and *FKBP4*, as pan-cancer biomarkers. *FKBP4* has already been discussed in a pan-cancer context, reinforcing its relevance across multiple cancer types[40]. Similarly, *DPM1* has been identified as a significant prognostic marker in hepatocellular carcinoma and breast cancer, highlighting its critical role in these individual cancers[39,45]. Our findings further support the consistent importance of *DPM1* across all 13 tasks, suggesting its potential as a pan-cancer biomarker. These genes could serve as key components in the development of diagnostic and therapeutic strategies applicable to a wide range of cancers, making them promising targets for further studies.

Similar to all other studies, our study has some limitations. Frist, although our scRNA-seq dataset included 13 cancer types, it did not encompass all possible cancer types, which could affect the generalizability of our findings. However, our approach demonstrated robust performance even in cancer types not included in the original scRNA-seq data, such as the grading task in PRAD, highlighting its broad applicability. Second, while we propose that our gene sets could enhance multi-modal model performance, this has not yet been validated. Recent studies have increasingly utilized multi-modal approaches that combine transcriptomic data with other multi-omics or pathology images in pan-cancer studies. Although we did not explore this in the current study, further research will include validating our gene sets in multi-modal frameworks to fully leverage their potential.

In conclusion, our study provides a robust framework for optimizing feature selection in cancer genomics by combining high-resolution scRNA-seq data with advanced analytical techniques. These findings have great potential for the improvement of predictive models in oncology and the advancement of personalized cancer treatment strategies. The identification of gene sets with significant pan-cancer relevance highlights the impact of our approach and paves the way for future research and clinical applications in cancer genomics.

In conclusion, our study provides a robust framework for optimizing feature selection in cancer genomics by combining high-resolution scRNA-seq data with advanced analytical techniques. Our findings have significant potential for improving predictive models in oncology, particularly in enhancing the performance of pan-cancer downstream tasks. By identifying gene sets with notable pan-cancer relevance, our approach paves the way for future research that can lead to more accurate and reliable cancer diagnostics and prognostics. Furthermore, our results offer valuable insights that can be applied to multi-modal frameworks, thereby advancing personalized cancer treatment strategies and contributing to the broader field of cancer genomics.

## Methods

### Datasets

To establish the gene expression landscape across a wide range of cancers, we obtained scRNA-seq data for 181 tumor biopsy samples from 13 different cancer types, which were collected by Nieto., et al[18]. We obtained RNA-seq data for 7,178 tumor samples across 16 cancer types from The Cancer Genome Atlas (TCGA), accessed via the NIH Genomic Data Commons Data Portal[46]. All datasets used in this study were downloaded using the GenomicDataCommons package in R.

To facilitate various downstream tasks, the types of cancer in this study have been grouped as follows. For the MSI classification, CRC, specifically colon adenocarcinoma (COAD) and rectal adenocarcinoma (READ), were collectively analyzed under a combined TCGA project designation. Similarly, for subtype classification, lung cancers were categorized into non-small cell lung cancer (NSCLC), combining both LUAD and LUSC. Kidney cancers were grouped into renal cell carcinoma (RCC), involving kidney renal clear cell carcinoma (KIRC), kidney renal papillary cell carcinoma (KIRP), and kidney chromophobe (KICH). Melanomas (MEL), which include skin cutaneous melanoma (SKCM) and uveal melanoma (UVM), were classified separately due to their different anatomical origins, despite both being melanomas. Lastly, gynecologic cancers (GYN) were grouped into ovarian serous cystadenocarcinoma (OV), uterine corpus endometrial carcinoma (UCEC), cervical squamous cell carcinoma, and endocervical adenocarcinoma (CESC).

### RNA-seq data pre-processing

RNA-seq data were pre-processed to ensure consistency across samples and to facilitate accurate comparative analysis. Read counts were first normalized by adjusting for gene length, which accounts for variations in transcript size. The normalized counts were then scaled to counts per million to standardize for sequencing depth, providing a comprehensive view of gene expression levels. Then, the read counts were log-transformed to stabilize variance, enhancing the analytical exploration of the gene expression landscape across cancer samples[47].

### Data preprocess for pan-cancer downstream tasks

Based on this processed data TMB scores were calculated for each sample by further normalizing read counts against gene lengths and adjusting for sequencing depth. First, the RNA-seq data were mapped to barcodes for each sample. Then, read counts were divided by gene lengths and were scaled by a factor of $10^6$ bases to derive TMB scores for each sample. Samples with a TMB score of higher than 10 were categorized as TMB-high, while those with a TMB score below 10 were classified as TMB-low[48].

To determine MSI scores, we analyzed the presence of missense mutations in key DNA mismatch repair (*MMR*) genes, including *MLH1, MLH3, MSH2, MSH3, MSH4, MSH5, MSH6, PMS1*, and *PMS2*[49]. Samples were classified as either microsatellite unstable or stable based on these mutations.

To classify prostate cancer samples for the grading task, we used gleason scores to classify them into three groups: low, moderate, and high. Clinical data was downloaded from cBioPortal, and samples were grouped based on their gleason scores[50].

Those with a score of 6 were placed in the low group, a score of 7 in the moderate group, and scores of 8, 9, and 10 in the high group[51].

For the mutation prediction tasks, we utilized somatic mutation datasets from TCGA to identify mutations in specific target genes across various cancer types. We examined each sample to determine whether the genes of interest were mutated or not. This mutation information was then cross-referenced with RNA-seq data to ensure consistency in the samples. Only those patients who had both mutation data and corresponding RNA-seq data were included in the final dataset for further analysis.

## Feature selection through high dimensional weighted gene co-expression network analysis

To identify co-expressed gene modules in the pan-cancer scRNA-seq dataset, we utilized the hdWGCNA package to perform weighted gene co-expression network analysis (WGCNA)[52]. First, metacells were constructed for each immune cell subset to optimize the dataset for the hdWGCNA pipeline. Principal Component Analysis (PCA) was used for dimensionality reduction, followed by the k-Nearest Neighbors algorithm to group similar cells for aggregation. The metacell gene expression matrix was then normalized and used as input for WGCNA. A suitable soft-thresholding power was selected to strengthen the correlations, and highly correlated modules were identified. Harmonized module eigengenes (hMEs) were calculated across the dataset based on these co-expression modules. For network visualization, we generated a UMAP embedding of the topological overlap map, which effectively illustrated the modular structure and relationships among the selected gene sets.

## Pathway enrichment analysis

Pathway enrichment analysis was performed using the GO Biological Process database[53]. Based on the top 100 genes selected from each module identified through hdWGCNA using kME scores, we conducted enrichment analysis for each module. The log-transformed P-value was used as the enrichment parameter. The top 5 enriched terms were selected for further analysis. Results were obtained via the R Enrichr package[54].

## Model development

We developed two separate model architectures optimized for RNA-seq data: a multi-layer perceptron (MLP) model and a graph neural network (GNN) model.

**MLP**

We developed an MLP model consisting of an embedding module and a classifier module. The embedding module reduces the high-dimensional input features into a lower-dimensional space through a single linear layer. This dimensionality reduction is facilitated by batch normalization and Leaky ReLU activation, which enhance the stability and improve the ability of the model to capture patterns in the data[55]. To mitigate overfitting, a dropout layer is incorporated. The classifier module processes these features through three sequential linear layers, each incorporating SELU activation and an additional dropout layer[56]. This configuration ensures that feature variability is conserved and increases robustness in classification tasks. Additionally, the MLP architecture is designed to adjust the number of neurons in each layer according to the specific needs of each downstream task.

**GNN**

We developed a GNN model based on a previous study[57]. The gene embedding module starts with an input embedding layer that maps high-dimensional input features into a hidden dimension using a linear layer, followed by dropout to prevent overfitting. The core of this module consists of multiple graph transformer layers that perform message passing and update node features, incorporating batch normalization and layer normalization to ensure stability and effective learning. Each graph transformer layer uses multi-head attention to capture diverse patterns in the data. The final output from the graph layers is reshaped and passed through a fully connected linear layer, followed by batch normalization and dropout. This embedding is then flattened to form the input to the classifier module. The classifier module processes the concatenated node embeddings through three sequential linear layers with SELU activation functions and dropout. Batch normalization is applied after each linear layer to maintain feature stability.

**Model training**

Both the MLP and GNN models were trained, employing binary cross-entropy for binary classification tasks and cross-entropy loss functions for multi-class classification tasks. Optimization was achieved using the Adam optimizer with a mini-batch size of 32 and a learning rate of 0.01[58]. Early stopping based on validation loss, with patience of 30 epochs, was used to prevent overfitting. To evaluate the model's performance, stratified cross-validation was performed[59]. Each fold was trained and validated, and the model's performance was further assessed through 100 bootstrap iterations on the test set to compute a 95% confidence interval for the results. This method ensures the model's reliability and accuracy across different patient samples. All computations were performed using PyTorch (version 2.0.1) on an NVIDIA A100 GPU (40 Gb) equipped with CUDA version 11.7.

**Feature importance through XGBoost**

In our study, XGBoost was used to evaluate feature importance from gene sets identified by hdWGCNA[60]. For each downstream task, we performed 100 bootstrap iterations to assess the importance of gene features. In each iteration, we resampled the dataset, trained the XGBoost classifier, and evaluated its performance using metrics such as AUROC and AUPRC. This rigorous bootstrap approach helped us to identify genes with consistently high feature importance, highlighting their importance in biological processes relevant to our study.

## Data availability

All scRNA-seq data utilized in this study are publicly accessible. The datasets were obtained from links provided by Nieto et al.[18] https://doi.org/10.5281/zenodo.4263972.

The remaining gene sets used for downstream tasks were also downloaded from links provided in their respective publications: Chen et al.[19], Mohammed et al.[20], Park et al.[21], Chen et al.[22], Liu et al.[23], Yuan et al.[24] Lastly, cancer-specific gene lists were obtained directly from the OncoKB[25] website https://www.oncokb.org/cancer-genes.

## Code availability

The source code and software pipeline to reproduce our study can be accessed at https://github.com/kimjh0107/2024$_p$ancancer$_s$cRNA.git.

# Acknowledgements

# Author information

**Authors and Affiliations LG AI Research, Seoul, South Korea** Jong Hyun Kim, Jongseong Jang

### Contributions

J.H.K. and J.J. conceived of and designed the study. J.H.K analyzed the data and performed data processing and visualization. J.J. supervised the project. J.H.K. and J.J. wrote the manuscript.

### Corresponding authors

Correspondence to Jongseong Jang.

# Ethics declarations

### Conflict of interest

The authors declare no competing interests.

# Additional information

| Task | Model | hdWGCNA 5-Fold Avg AUROC | hdWGCNA XGBOOST 5-Fold Avg AUROC | Chen et al.[19] 5-Fold Avg AUROC | Mohammed et al.[20] 5-Fold Avg AUROC | Park et al.[21] 5-Fold Avg AUROC | Chen et al.[22] 5-Fold Avg AUROC | Liu et al.[23] 5-Fold Avg AUROC | Yuan et al.[24] 5-Fold Avg AUROC | OncoKB[25] 5-Fold Avg AUROC |
|---|---|---|---|---|---|---|---|---|---|---|
| TMB-LUAD | MLP | 0.757 (0.747-0.766) | **0.791 (0.782-0.800)** | 0.752 (0.743-0.762) | 0.710 (0.700-0.721) | 0.720 (0.710-0.730) | 0.739 (0.729-0.749) | 0.627 (0.616-0.639) | 0.586 (0.573-0.597) | 0.746 (0.737-0.756) |
|  | GNN | 0.744 (0.735-0.754) | **0.764 (0.754-0.773)** | 0.737 (0.727-0.747) | 0.721 (0.711-0.731) | 0.741 (0.731-0.750) | 0.737 (0.727-0.747) | 0.668 (0.657-0.678) | 0.541 (0.496-0.524) | 0.730 (0.720-0.740) |
| TMB-LUSC | MLP | 0.538 (0.526-0.550) | **0.647 (0.636-0.659)** | 0.548 (0.537-0.559) | 0.558 (0.545-0.571) | 0.528 (0.516-0.540) | 0.536 (0.524-0.548) | 0.542 (0.530-0.553) | 0.543 (0.532-0.555) | 0.555 (0.543-0.567) |
|  | GNN | 0.538 (0.526-0.550) | **0.604 (0.592-0.616)** | 0.575 (0.564-0.587) | 0.569 (0.557-0.581) | 0.495 (0.483-0.506) | 0.588 (0.576-0.600) | 0.481 (0.470-0.493) | 0.546 (0.534-0.558) | 0.551 (0.539-0.562) |
| TMB-SKCM | MLP | 0.720 (0.709-0.731) | **0.772 (0.762-0.782)** | 0.759 (0.749-0.769) | 0.699 (0.688-0.709) | 0.644 (0.632-0.655) | 0.652 (0.641-0.664) | 0.661 (0.649-0.673) | 0.550 (0.538-0.562) | 0.745 (0.735-0.755) |
|  | GNN | 0.632 (0.619-0.644) | **0.747 (0.736-0.758)** | 0.634 (0.623-0.645) | 0.659 (0.647-0.671) | 0.597 (0.586-0.609) | 0.681 (0.670-0.692) | 0.610 (0.599-0.622) | 0.588 (0.578-0.597) | 0.628 (0.616-0.640) |
| TMB-CRC | MLP | 0.776 (0.766-0.786) | **0.826 (0.818-0.835)** | 0.800 (0.790-0.809) | 0.783 (0.773-0.793) | 0.759 (0.749-0.770) | 0.797 (0.788-0.806) | 0.749 (0.739-0.760) | 0.747 (0.736-0.758) | 0.787 (0.778-0.797) |
|  | GNN | 0.763 (0.753-0.774) | **0.808 (0.799-0.817)** | 0.781 (0.771-0.790) | 0.762 (0.752-0.773) | 0.708 (0.697-0.719) | 0.780 (0.770-0.790) | 0.776 (0.766-0.786) | 0.701 (0.690-0.712) | 0.755 (0.744-0.765) |
| MSI-STAD | MLP | 0.971 (0.967-0.976) | **0.990 (0.988-0.991)** | 0.982 (0.979-0.985) | 0.958 (0.952-0.964) | 0.911 (0.904-0.918) | 0.982 (0.979-0.984) | 0.914 (0.905-0.922) | 0.879 (0.870-0.887) | 0.977 (0.974-0.979) |
|  | GNN | 0.962 (0.957-0.967) | 0.979 (0.975-0.983) | 0.945 (0.940-0.950) | 0.935 (0.928-0.941) | 0.801 (0.788-0.813) | 0.976 (0.972-0.979) | 0.869 (0.860-0.879) | 0.849 (0.839-0.859) | **0.982 (0.979-0.985)** |
| MSI-CRC | MLP | 0.916 (0.909-0.923) | **0.931 (0.925-0.937)** | 0.915 (0.907-0.923) | 0.917 (0.909-0.925) | 0.874 (0.864-0.884) | 0.898 (0.890-0.906) | 0.911 (0.903-0.918) | 0.858 (0.850-0.866) | 0.925 (0.918-0.932) |
|  | GNN | 0.897 (0.889-0.906) | **0.936 (0.930-0.942)** | 0.910 (0.902-0.918) | 0.915 (0.908-0.922) | 0.841 (0.830-0.851) | 0.908 (0.899-0.916) | 0.917 (0.910-0.924) | 0.828 (0.818-0.838) | 0.893 (0.885-0.902) |
| MUT-BRCA-TP53 | MLP | 0.931 (0.927-0.934) | 0.945 (0.942-0.948) | 0.942 (0.939-0.946) | 0.888 (0.883-0.893) | 0.919 (0.915-0.923) | 0.921 (0.917-0.925) | 0.871 (0.865-0.876) | 0.811 (0.804-0.817) | **0.951 (0.949-0.954)** |
|  | GNN | 0.925 (0.921-0.929) | 0.934 (0.930-0.942) | 0.930 (0.926-0.934) | 0.884 (0.879-0.889) | 0.893 (0.888-0.898) | 0.921 (0.917-0.925) | 0.869 (0.864-0.875) | 0.773 (0.769-0.778) | **0.935 (0.932-0.938)** |
| MUT-LUAD-TP53 | MLP | 0.862 (0.854-0.869) | **0.869 (0.862-0.876)** | 0.861 (0.853-0.869) | 0.756 (0.746-0.766) | 0.772 (0.762-0.782) | 0.854 (0.846-0.861) | 0.722 (0.712-0.732) | 0.663 (0.652-0.674) | 0.857 (0.849-0.865) |
|  | GNN | 0.848 (0.840-0.856) | **0.872 (0.866-0.880)** | 0.842 (0.834-0.849) | 0.773 (0.763-0.783) | 0.791 (0.781-0.800) | 0.838 (0.830-0.846) | 0.663 (0.655-0.671) | 0.615 (0.608-0.621) | 0.844 (0.835-0.852) |
| MUT-STAD-TP53 | MLP | 0.751 (0.740-0.761) | **0.790 (0.779-0.800)** | 0.766 (0.756-0.776) | 0.717 (0.706-0.728) | 0.642 (0.630-0.654) | 0.734 (0.723-0.744) | 0.647 (0.634-0.659) | 0.625 (0.612-0.638) | 0.764 (0.754-0.774) |
|  | GNN | 0.723 (0.712-0.733) | **0.790 (0.780-0.799)** | 0.713 (0.702-0.724) | 0.681 (0.668-0.693) | 0.592 (0.579-0.604) | 0.703 (0.692-0.714) | 0.619 (0.606-0.632) | 0.593 (0.580-0.606) | 0.680 (0.669-0.691) |
| MUT-LUAD-EGFR | MLP | 0.817 (0.804-0.830) | **0.868 (0.858-0.878)** | 0.834 (0.821-0.846) | 0.803 (0.791-0.815) | 0.797 (0.786-0.808) | 0.814 (0.801-0.826) | 0.721 (0.706-0.737) | 0.687 (0.672-0.701) | 0.834 (0.823-0.846) |
|  | GNN | 0.757 (0.744-0.770) | **0.865 (0.855-0.875)** | 0.735 (0.722-0.748) | 0.786 (0.773-0.798) | 0.705 (0.691-0.719) | 0.793 (0.781-0.805) | 0.646 (0.630-0.661) | 0.677 (0.663-0.691) | 0.801 (0.788-0.813) |
| MUT-LUAD-KRAS | MLP | 0.793 (0.782-0.803) | **0.845 (0.835-0.854)** | 0.824 (0.815-0.832) | 0.754 (0.743-0.764) | 0.640 (0.628-0.651) | 0.777 (0.766-0.788) | 0.703 (0.692-0.715) | 0.614 (0.600-0.627) | 0.806 (0.797-0.815) |
|  | GNN | 0.743 (0.732-0.754) | **0.792 (0.782-0.801)** | 0.762 (0.751-0.773) | 0.699 (0.751-0.773) | 0.678 (0.666-0.689) | 0.737 (0.727-0.748) | 0.607 (0.594-0.619) | 0.610 (0.597-0.623) | 0.698 (0.687-0.710) |
| MUT-PAAD-KRAS | MLP | 0.858 (0.845-0.871) | **0.904 (0.894-0.915)** | 0.851 (0.838-0.864) | 0.853 (0.838-0.867) | 0.744 (0.728-0.760) | 0.845 (0.831-0.859) | 0.824 (0.810-0.838) | 0.797 (0.781-0.812) | 0.884 (0.873-0.896) |
|  | GNN | 0.853 (0.840-0.866) | **0.892 (0.879-0.904)** | 0.871 (0.859-0.883) | 0.846 (0.833-0.860) | 0.771 (0.757-0.786) | 0.868 (0.856-0.881) | 0.795 (0.780-0.809) | 0.801 (0.787-0.815) | 0.875 (0.864-0.886) |
| GRAD-PRAD | MLP | 0.731 (0.721-0.740) | **0.779 (0.771-0.787)** | 0.752 (0.743-0.761) | 0.686 (0.676-0.696) | 0.663 (0.654-0.673) | 0.726 (0.717-0.736) | 0.692 (0.683-0.702) | 0.519 (0.510-0.529) | 0.748 (0.739-0.757) |
|  | GNN | 0.720 (0.712-0.729) | 0.695 (0.686-0.704) | 0.710 (0.701-0.719) | 0.616 (0.606-0.627) | 0.684 (0.675-0.693) | 0.745 (0.737-0.753) | 0.632 (0.622-0.641) | 0.558 (0.548-0.568) | **0.754 (0.746-0.762)** |

**Table 1. Comparative performance analysis of gene sets for downstream task.** We report the downstream classification task performance on two different deep learning frameworks, MLP and GNN. Tasks include TMB assessment in LUAD, LUSC, SKCM, and CRC; MSI classification in STAD and CRC; and mutation prediction for genes (*TP53*, *EGFR*, and *KRAS*) across BRCA, LUAD, and STAD. Results are based on 5-fold validation, with values representing the average AUROC and 95% confidence intervals (CI) values obtained via bootstrapping. The numbers in parentheses indicate the lower and upper bounds of the 95% CI. The best performance for each task is in bold.

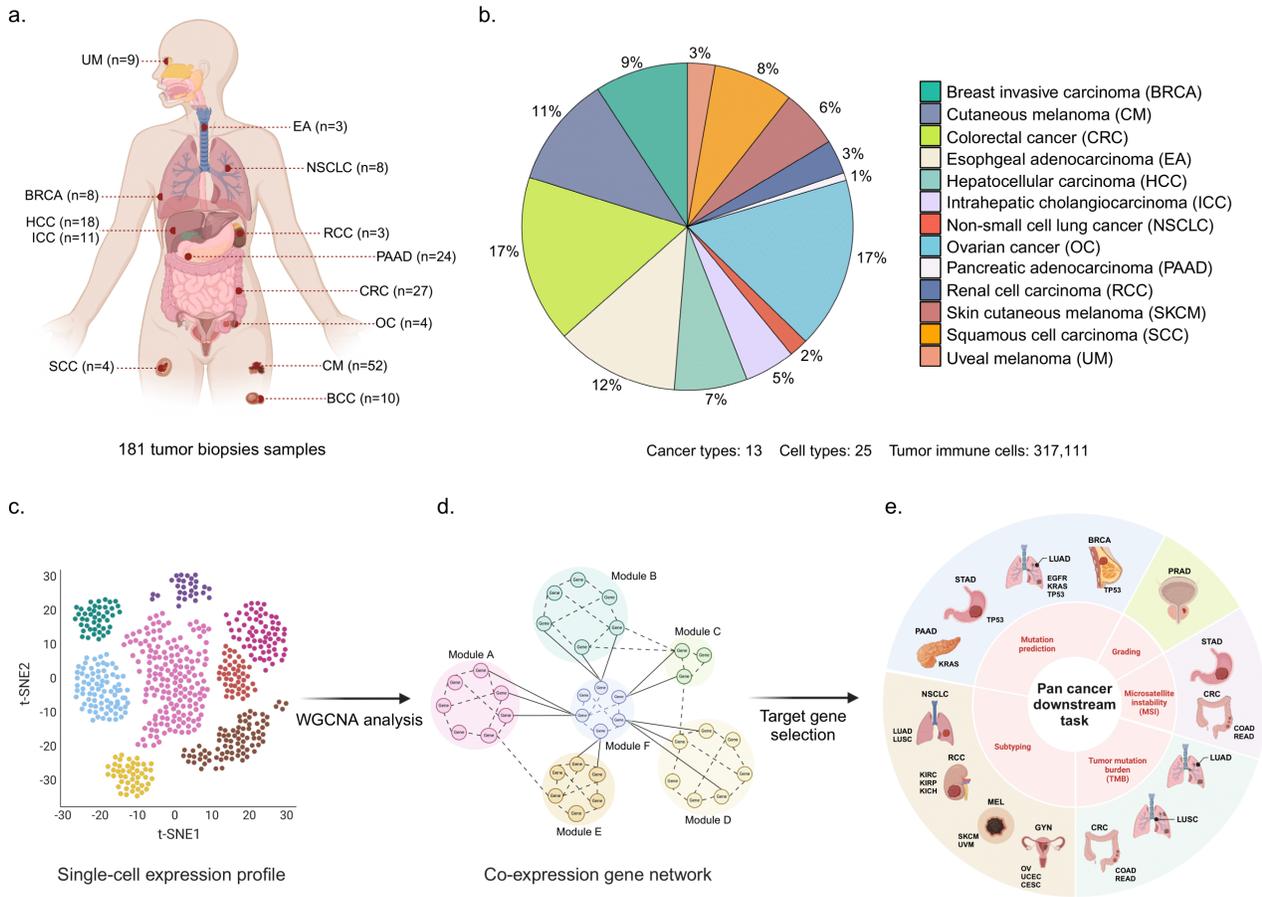

**Fig. 1. Schematic overview of the study workflow. a.** Each cancer type is indicated along with the number of samples analyzed: Uveal melanoma (UM, n=9), esophageal adenocarcinoma (EA, n=3), non-small cell lung cancer (NSCLC, n=8), breast invasive carcinoma (BRCA, n=8), hepatocellular carcinoma (HCC, n=18), intrahepatic cholangiocarcinoma (ICC, n=11), renal cell carcinoma (RCC, n=3), pancreatic adenocarcinoma (PAAD, n=24), colorectal cancer (CRC, n=27), ovarian cancer (OC, n=4), squamous cell carcinoma (SCC, n=4), cutaneous melanoma (CM, n=52), basal cell carcinoma (BCC, n=10). **b.** Proportional distribution of the 13 cancer types within the scRNA-seq dataset, highlighting the diversity and representation of different tumor types used in the study. The dataset includes 317,111 tumor immune cells classified into 25 distinct cell types. **c.** Schematic UMAP visualization of single-cell expression profiles, which were prepared for hdWGCNA. **d.** Schematic of the co-expression gene network derived from hdWGCNA. The genes comprising this biological module were used for various downstream analyses through the target gene process. **e.** (e) Application of selected co-expressed gene modules to various pan-cancer downstream tasks. These tasks include mutation prediction, cancer subtyping, tumor mutation burden (TMB) assessment, microsatellite instability (MSI) classification, cancer subtyping, and grading.

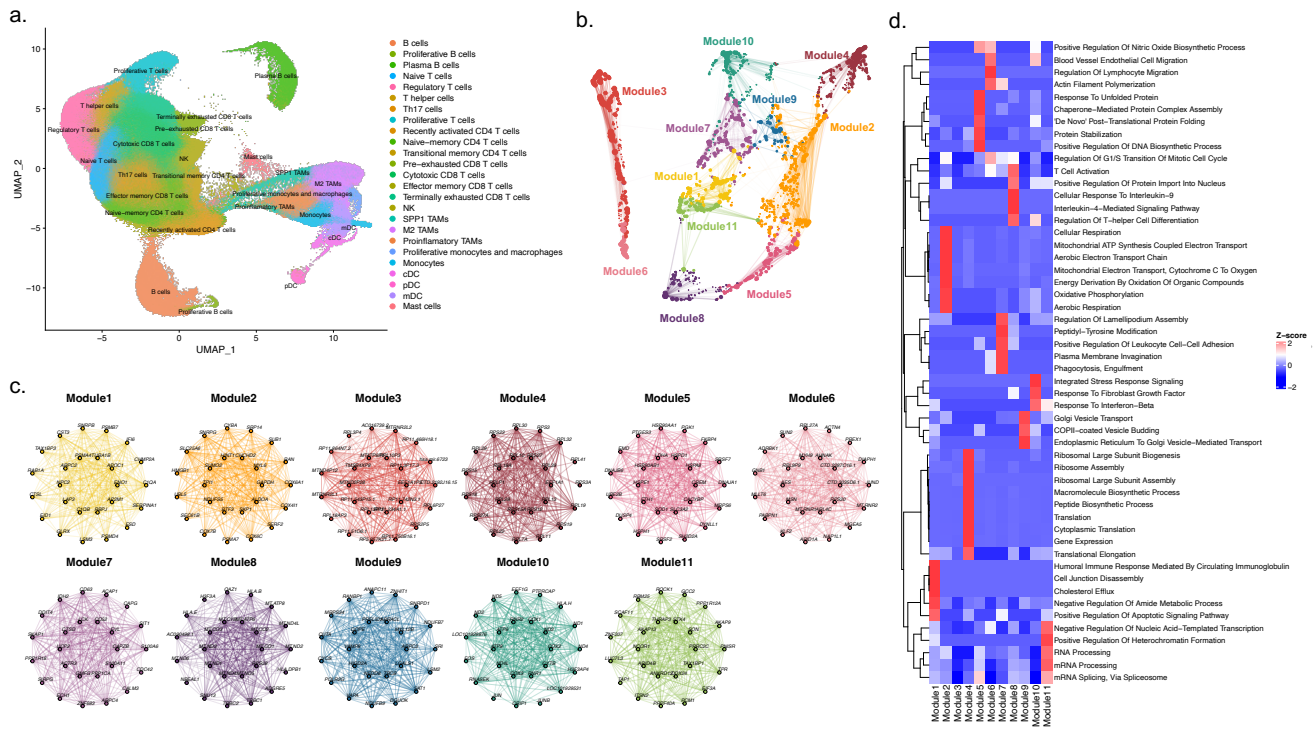

**Fig. 2. Comprehensive gene co-expression networks reveal key functional modules in the tumor immune microenvironment. a.** UMAP embeddings display the distribution of 25 immune cell types across 181 tumor biopsy samples, visualizing all 317,111 cells. Each color represents a distinct immune cell type. **b.** UMAP representation of the gene co-expression network. A total of 11 modules were identified. Nodes indicate individual genes, and edges signify co-expression relationships between genes and hub genes within modules. Node sizes are proportional to their kME (eigengene-based connectivity) values. Colors denote different co-expression modules. **c.** Visualisation of hub gene networks for each spatial co-expression module. The 25 highest-ranked hub genes based on kME are presented. In the network, nodes represent genes, while edges indicate co-expression links. **d.** Heatmap summarizing GO pathway enrichment analysis for each module. Each row represents a biological process, and columns correspond to modules, with the color scale indicating Z-score values.

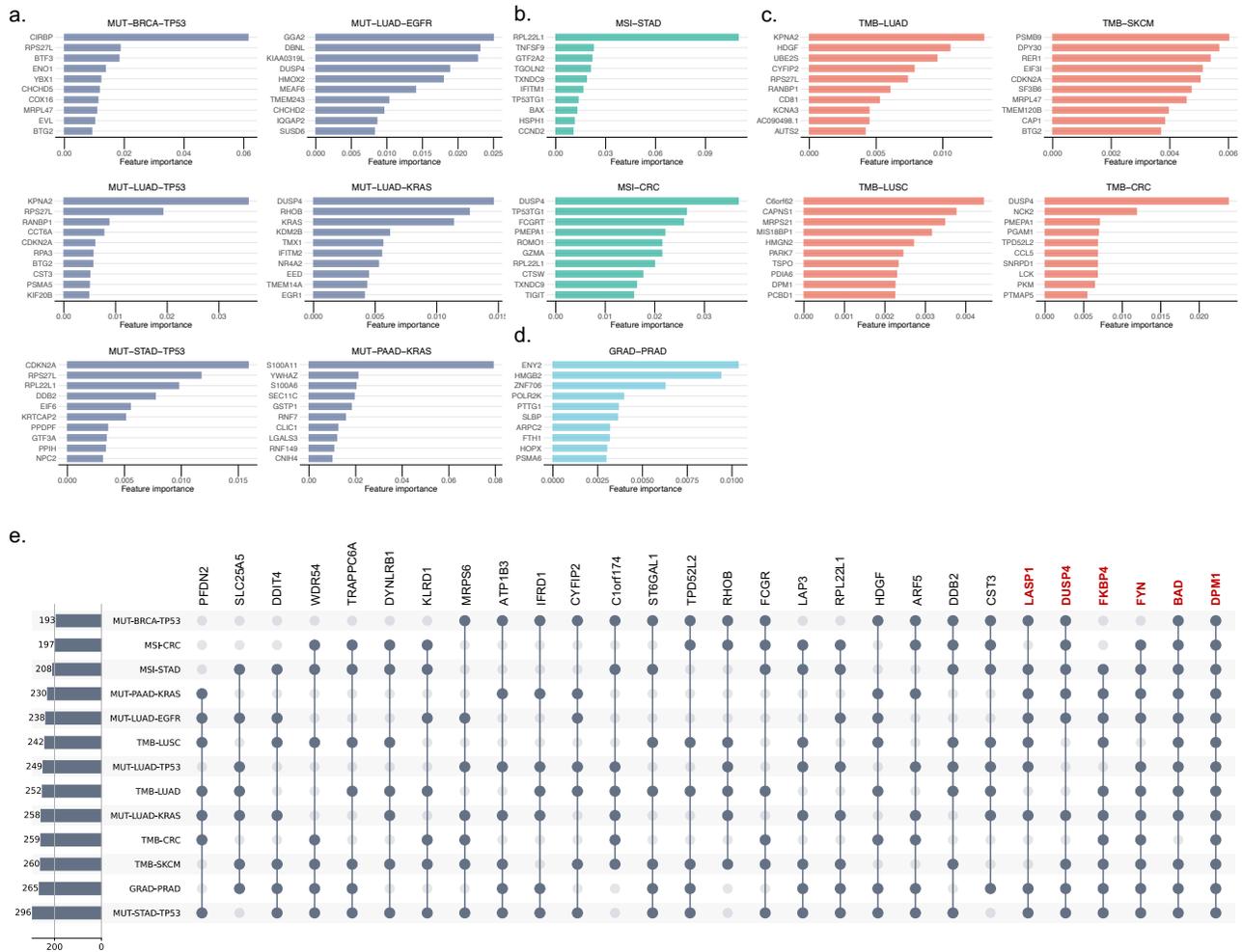

**Fig. 3. Feature importance analysis and overlap of key genes across downstream tasks. a-d.** Feature importance scores for the top 10 genes were extracted for each downstream task, with figures representing: **a.** mutation prediction (MUT), **b.** microsatellite instability (MSI), **c.** tumor mutation burden (TMB), and **d.** cancer grading (GRAD). Each panel displays the importance scores for the respective genes, highlighting those with the highest relevance to each task. **e.** The upset plot illustrates the overlap of key genes among the different downstream tasks. Genes consistently ranking high in feature importance across multiple tasks are shown, with red-bolded genes indicating the most significant overlaps. The left bar plot presents the number of features selected for each task.